# SPIM Architecture for MVC based Web Applications

**R. Sridaran**
Professor & HOD, Department of MCA, New Horizon College of Engineering, Bangalore-560087.
Email: sridaran.rajagopal@gmail.com

**G. Padmavathi**
Professor & Head, Department of Computer Science, Avinashilingam University for Women, Coimbatore-641047.
Email: ganapathii.padmavathi@gmail.com

**K. Iyakutti**
Senior Professor, School of Physics, Madurai Kamaraj University, Madurai-625021.

**M.N.S. Mani**
Consultant, Lakshmi Systems, Madurai-625020.

--------------------------------------------------ABSTRACT-------------------------------------------------
The Model / View / Controller design pattern divides an application environment into three components to handle the user-interactions, computations and output respectively.  This separation greatly favors architectural reusability. The pattern works well in the case of single-address space and not proven to be efficient for web applications involving multiple address spaces.  Web applications force the designers to decide which of the components of the pattern are to be partitioned between the server and client(s) before the design phase commences.  For any rapidly growing web application, it is very difficult to incorporate future changes in policies related to partitioning. One solution to this problem is to duplicate the Model and controller components at both server and client(s). However, this may add further problems like delayed data fetch, security and scalability issues.  In order to overcome this, a new architecture SPIM has been proposed that deals with the partitioning problem in an alternative way.  SPIM shows tremendous improvements in performance when compared with a similar architecture.

## I. INTRODUCTION

Design Patterns [1] are extensions of object-oriented programming and hence promote reusability. Pattern structures include more of interfaces than implementations.  Therefore, it can be plugged into any application to address a specific issue without affecting the remaining part of code or functionality.  The Model / View / Controller (MVC) [2] design pattern is generally applied to simplify the architectural design of an application.  In the case of web applications, the MVC may suffer from partitioning issues as narrated under section 1.3.

### 1.1 THE MODEL / VIEW / CONTROLLER DESIGN PATTERN

The MVC design pattern suggests the division of a problem into three parts as follows: The Model is used to hold the computational parts of the program; the View is to deal with the rendering of output and the Controller is to govern the interaction between the View and the user.  This classification has been widely accepted since it promotes architectural reusability.  The Model component contains the core data and functionality.  This is independent of specific output representations or input behavior.  Controllers receive inputs in the form of events that are translated into service requests for the Model or the View. All user interactions with the system will be only through the Controllers.

The MVC approach has many advantages.
- Multiple Views of the same Model can be used simultaneously. New data Views can also be introduced at any point of time.
- MVC prevents tight coupling between the objects. Since the dependency of a class with so many other classes is reduced, the class can easily be re-used.
- The application's look and feel can be altered without affecting the business logic or the data.
- Different interfaces or user levels can be maintained by the same application.
- The entire application can be built or managed independently by business logic developers, flow of control developers and web page designers.
- The MVC is ideal to maintain an environment comprising of different technologies across different locations.
- MVC promotes scalability and maintainability.

### 1.2 MVC FOR WEB APPLICATIONS

Web applications have to be partitioned between the client and the server. It is always appropriate to keep the view at the client machine as it deals only with the rendering of output against a client's request. Similarly, it is ideal to keep the Model at the server side as it contains the business logic. The controller can be kept at the server or at the client side depending upon the application requirement. Since, the Model, view and controller frequently interact with each other they are designed to be kept in a single address space. In contrast to this, web-based applications supporting multiple views with different kind of user requests essentially occupy multiple address spaces and hence do not serve as the right platform for this pattern usage [3].

Using MVC for Internet applications face few more challenges when compared with intranet based applications. Since uncontrolled amount of users will be interacting with the application, it is essential to address several issues. The Internet applications should guarantee the information availability at an economical price. A wide range of customers spread across the globe will be ever growing. So scalability issues should also be addressed. With limited resources being shared by unlimited users, it should also reinforce security.

### 1.3 PARTITIONING ISSUES

The problem with using the MVC design pattern to develop web applications arises from the fact that web applications are intrinsically partitioned between the client and the server. The view is displayed on the client; but the Model and controller can be partitioned in many ways between the client and server. The developer is forced to partition the web application during the design phase itself. In contrast to this, MVC is partition-independent. In other words, the Model, view and controller reside and execute in a single address space where the partitioning issues do not arise. Partition independence is the main feature of MVC since location dependency should not drive architecture decisions. Unfortunately partitioning implies that web applications are location dependent. Hence it is much difficult to apply the MVC design patterns in the web application scenario.

Many a times, it is not possible to make correct partitioning decisions during the design phase itself since these decisions depend upon the application requirements that may change from time to time. The correct partitioning decision also depends upon certain static factors like system architectures of client and server and dynamic factors like network congestion. To summarize, web applications can use the MVC pattern when the correct partitioning is known and the available technology infrastructure is compatible with the partitioning.

The problem of partitioning prevents the MVC pattern being used effectively by web application designers. The new architecture proposed, SPIM (Sridaran-Padmavathi-Iyakutti-Mani) as explained under Section 3, provides a flexible approach of applying MVC for web applications.

### 1.4 INTRODUCTION TO MASHUP

Mashup is a web application that combines data from more than one source into a single integrated tool. The term 'Mashup' implies easy, fast integration, frequently done by access to open APIs and data sources to produce results that are not the original goal of the data owners. It may also be regarded as a web page or application that integrates complementary elements from two or more sources. The data from more than one source into a single integrated tool have been used for this purpose. A mashup application has got three parts:

- A web page that provides a new service using its own data and data from other sources
- Additional content provider to make data available across the web through an API and using different web protocols or other web services.
- The client, the user of the mashup, often using a web browser, displaying a web page, containing the mashup.

### 1.5 ORGANIZATION OF THE PAPER

Section 2 provides a short survey of web applications employing MVC pattern. Section 3 provides the implementation details of SPIM architecture. The proposed architecture SPIM is analyzed with a similar architecture dmvc [3] in Section 4.

## 2. RELATED WORK

Bodhuin T et al. [4] present a strategy by using MVC for migrating a legacy COBOL system into a web-enabled architecture. The needed information from the COBOL source code is extracted and then wrapper classes are applied to convert them into JSP.

Yu Ping et al. [5] have provided a methodology by which the database functionalities are extracted from the source programs of a legacy web application to form JavaBean objects. The legacy presentation components are translated into JSP pages that are made to refer the JavaBean objects and the linkage information are extracted and web site is made Controller-centric. This work is being extended to incorporate the language independent feature in the source program of the translation process.

MVC has been successfully applied to numerous web applications. An educational tool has been proposed by Vichido. C et al. [6] in which a user-modeling server has been enhanced to enable a web-based learning environment. In the same education scenario, the management of a research projects has been dealt by Liyong Zhang et al. [7] where MVC has been employed to cater to the information requirements. In another educational tool proposed by da Sliva et al. [8] MVC has been used to allow functionality extensions.

In the case of database oriented web applications, the application introduced by Selfa DM. et al. [9] employs MVC in different phases of analysis, design and implementation with a central database made up of multiple relations and a large number of web pages. In the same manner, an architecture of web-oriented warfare has been proposed by Xiaofei Wang et al. [10] that has applied MVC pattern for the implementation of an interface agent.

Xiaohong Qiu [11] has applied a message-based MVC pattern for the construction of desktop applications to incorporate web services. In this application, the pattern has been used to provide the web services that are accessible from different client platforms.

## 3. ARCHITECTURE OF SPIM

The structure of SPIM architecture is shown in Fig 1. The architecture consists of a Model-Controller pair one at each side of server and client respectively. The Server Model (SM) and Server Controller (SC) are kept at server side and the Client Model (CM) is kept at client side. The Client Controller (CC) is only visible to the users of the system. The Data Store (DS) kept at server side is accessible by SM. Similarly the Data Cache (DC) that holds the most recently used segments of data is accessible by CM. All requests are passed in the form of XML communications. DC can be thought of an intermediate storage used for storing the results of the user requests.

The CV, responsible for rendering of output can also be part of another application as shown in Fig 1. In other words, CV may belong to a 'third party' willing to share the services of SPIM. Here the middle tier fetches the necessary services from the server, as expected by the client. The third party application mentioned over here, can take up other services also. The same results rendered by the architecture may be displayed with different views as required by the target third party applications thereby adding flexibility for the users of the system. The architecture also facilitates mixing of different combination of services to 'third-party' views which makes it as a 'Mash-up' application.

The sequence of steps involved in the request-response

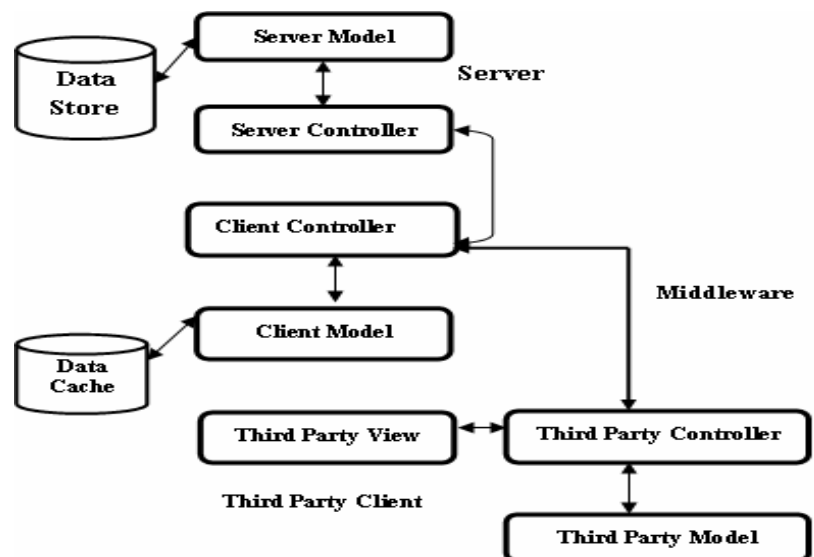

Figure 1: Architecture of SPIM

### 3.1 THE REQUEST-RESPONSE PROCESS OF SPIM

process of SPIM is given under:

Step 01: Request from CV (users) reaches CC
Step 02: CC passes the request to CM.
Step 03: CM fetches the data from Cache if available, else returns "ERROR" to CC.

Step 04: If data is available, the results are sent to CC. If "ERROR" returned, Go to Step 08.
Step 05: CC requests CM to generate the XML document of the result
Step 06: CM generates and passes the XML document to CC, which in turn send the same to CV.
Step 07: CV generates View and renders to the end user and End of Transaction.
Step 08: CC transfers the request to SC.
Step 09: SC in turn requests SM.
Step 10: SM fetches the required data from Data Store if available, else ERROR is returned to SC.
Step 11: SC sends the data to CC
Step 12: CC requests CM to store the resultant Data in DC and to generate the XML document of the result
Step 13: CM generates and passes the XML document to CC apart from storing the same in DC.
Step 14: CC in turn sends the resultant information to CV.
Step 15: CV generates View and renders to the end user and End of Transaction.

SPIM shields SC being accessed by unauthorized service requests thereby promoting security. The architecture further avoids the need for model synchronization since it is not permitting duplication of the same.

### 3.2 ASSUMPTIONS

The implementation of SPIM has been carried out with Java Server Pages (JSP) for the View, Servlets for Controllers and Enterprise Java Beans (EJB) for Models. My-SQL is the database used as DS and the development is carried out in Net Beans environment. DC is implemented as a text file.

### 4. ANALYSIS AND INTERPRETATION

The performance testing is carried out in both SPIM and dmvc by varying the record size of the database from 1000 to 30000 at regular intervals of 1000. The tests are carried out for data fetch from a large-scale database. The experiment is conducted as two cases a) For data fetch in DC and b) For data fetch in DS. The values of the experimentation are provided in TABLE 1. From the tabulated results, it is evident that SPIM shows improvements with respect to the time complexity measures.

Table 1: Test Results dmvc Vs SPIM

| Data Fetch from DC | | | | Data Fetch from DS | | | |
|---|---|---|---|---|---|---|---|
| Records | dmvc (t1) | SPIM (t2) | Decrease % | Records | dmvc (t1) | SPIM (t2) | Decrease % |
| 1000 | 15 | 15 | | 1000 | 32 | 16 | 50.00 |
| 2000 | 63 | 63 | | 2000 | 141 | 62 | 56.03 |
| 3000 | 156 | 140 | 10.26 | 3000 | 266 | 125 | 53.01 |
| 4000 | 219 | 218 | 0.46 | 4000 | 453 | 219 | 51.66 |
| 5000 | 688 | 672 | 2.33 | 5000 | 781 | 343 | 56.08 |
| 6000 | 500 | 492 | 1.60 | 6000 | 1047 | 516 | 50.72 |
| 7000 | 687 | 703 | 2.33 | 7000 | 1407 | 750 | 46.70 |
| 8000 | 1125 | 954 | 15.20 | 8000 | 1844 | 906 | 50.87 |
| 9000 | 1203 | 1216 | 1.08 | 9000 | 2515 | 1297 | 48.43 |
| 10000 | 3344 | 3260 | 2.51 | 10000 | 5844 | 2438 | 58.28 |
| 11000 | 4313 | 4282 | 0.72 | 11000 | 8515 | 4453 | 47.70 |
| 12000 | 5047 | 5212 | 3.27 | 12000 | 10422 | 5500 | 47.23 |
| 13000 | 6141 | 6219 | 1.27 | 13000 | 12391 | 6235 | 49.68 |
| 14000 | 8719 | 8744 | 0.29 | 14000 | 17438 | 8719 | 50.00 |
| 15000 | 10609 | 10469 | 1.32 | 15000 | 21359 | 10578 | 50.48 |
| 16000 | 12375 | 12563 | 1.52 | 16000 | 24672 | 12547 | 49.14 |
| 17000 | 14515 | 14406 | 0.75 | 17000 | 28188 | 14375 | 49.00 |
| 18000 | 17204 | 15985 | 7.09 | 18000 | 34515 | 16016 | 53.60 |
| 19000 | 19844 | 19375 | 2.36 | 19000 | 39281 | 18359 | 53.26 |
| 20000 | 23484 | 23287 | 0.84 | 20000 | 47141 | 23656 | 49.82 |
| 21000 | 25266 | 25750 | 1.92 | 21000 | 50438 | 25297 | 49.85 |
| 22000 | 28296 | 28313 | 0.06 | 22000 | 57282 | 28453 | 50.33 |
| 23000 | 31218 | 31469 | 0.80 | 23000 | 63312 | 31437 | 50.35 |

| 24000 | 34297 | 35032 | 2.14 | 24000 | 68532 | 35016 | 48.91 |
|---|---|---|---|---|---|---|---|
| 25000 | 40969 | 40363 | 1.48 | 25000 | 81625 | 40656 | 50.19 |
| 26000 | 42625 | 42219 | 0.95 | 26000 | 84422 | 42281 | 49.92 |
| 27000 | 46047 | 46484 | 0.95 | 27000 | 91890 | 46531 | 49.36 |
| 28000 | 50610 | 50266 | 0.68 | 28000 | 101531 | 49969 | 50.78 |
| 29000 | 54140 | 54375 | 0.43 | 29000 | 109750 | 54156 | 50.66 |
| 30000 | 62312 | 60797 | 2.43 | 30000 | 122079 | 61250 | 49.83 |
| | | Average | 1.25 | | | | 50.73 |

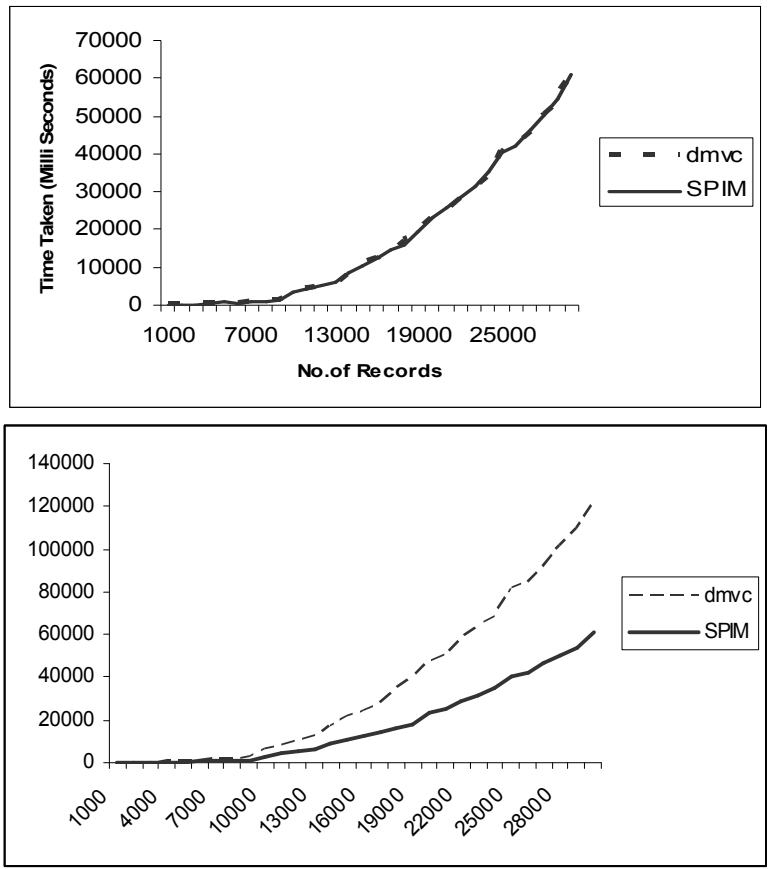

Figure 2: Graphs showing Time complexity measures for (a) Data Fetch in Cache and (b) in Data Store.

The graph patterns as shown in Fig 2 indicate the decrease in response times as the invocations grow.

The data will be found in DC in case it has been already fetched by a previously executed query. The implementations of dmvc and SPIM showed only an average reduction in time as 1.25% which is not very significant against the range of 1-30,000 record sets. However, the scenario changes in the case of newly executed queries where the required data may not be available in cache. Since SPIM does not allow duplication of models and controllers as in dmvc, the unnecessary overheads involved in searching the cache by SM and in searching the DS by CM are avoided. Due to this, the test cases where the data fetch is from DS showed an average reduction in time as 50.73%. This is expected to be more for multimedia data involving huge memory.

The Standard Deviations of time values of t1 and t2 for both the cases are calculated using the formula:

**Standard Deviation ($\sigma$) =** $\sqrt{\dfrac{\sum(x-\bar{x})^2}{(n-1)}}$

where x is the sample mean and n is the sample size. The following table, TABLE 2 shows the computed values of $\sigma$ for the two cases discussed above.

Table 2: Computed values of $\sigma$ for Cases A and B

| Architecture | Case A | Case B |
|---|---|---|
| **dmvc** | 19058.11 | *38049.14* |
| **SPIM** | *18952.33* | *18988.91* |

The σ values for SPIM across the two cases are found to be very consistent showing the steadiness of the proposed architecture.

## 5. CONCLUSION AND FUTURE WORK

The proposed SPIM architecture will definitely be helpful for the web application developers to make use of MVC pattern effectively without becoming entangled into the partitioning problem.  Using SPIM, the components of the web application can easily be managed independently.  It is also analyzed how SPIM is consistent and secured for web applications involving multiple services.  The future work involves development of an algorithm for migration of a legacy application into SPIM architecture.

## REFERENCES

[1] Erich Gamma, Richard Helm, Ralph Johnson, and John Vlissides, Design Patterns: Elements of Reusable Object-Oriented Software (Pearson Education, New Delhi).

[2] Frank Buschmann, Regine Meunier, Hans Rohnert, Peter Sommerlad, and Michael Stal Pattern-Oriented Software Architecture (John Wiley & Sons (Asia) Pte Ltd, 2001).

[3] Avraham Leff and James T. Rayfield, Web-Application Development Using the Model View Controller Design Pattern,  Proc. 5th IEEE Conf. on Enterprise Distributed Object Computing, 2001, 118-127.

[4] Bodhuin T. Guardabascio E.  and Tortorella M., Migrating COBOL systems to the Web by using the MVC design pattern, Proc. 9th Working Conf. on Reverse Engineering, 2002, 329-338.

[5] Yu Ping, Kostas Kontogiannis, and terrence C. Lau, Transforming Legacy Web Applications to the MVC Architecture, *Proc. 11th Annual International Workshop on Software Technology and Engineering Practice,* IEEE, 2003,133-142.

[6] Vichido, C. Estrada, M. Sanchez, A., A constructivist educational tool: software architecture for Web-based video games, *Proc. 4th Mexican International Conf. on Computer Science*, 2003,144 – 150.

[7] Liyong Zhang, Futing Ma, Chongquan Zhong, Li Zhang and Ying Wang, A MVCD Model and Its Application in University Project Management, *Proc. 6th World Congress on Intelligent Control and Automation*, 2006, Vol: 2,7008 – 7012.

[8] da Silva, E.Q. and de Abreu Moreira, D. WebMODE: a framework for development of Web-based tools for management of educational activities, *Proc. 5th IEEE International Conf. on Advanced Learning Technologies,* ICALT *2005*, July 2005, 922 – 924.

[9] Selfa D.M. Carrillo M. and Del Rocio Boone M., A Database and Web Application Based on MVC Architecture, *Proc. 16th International Conf. on Electronics, Communications and Computers*, IEEE Computer Society, 2006, 48.

[10] Xiaofei Wang, Yunqiu Chen, and Yuliang Liu, Web-oriented warfare command decision support system based on agent and data warehouse, Proc. International Conf. on Cyberworlds, 2005, 498.

[11] Xiaohong Qiu, Building desktop applications with Web services in a message-based MVC paradigm, *Proc. IEEE International Conf. on Web* Services, 2004, 765–768.

**Authors Biography**

**R. Sridaran** has obtained his post-graduation  in computer applications from Madurai Kamaraj University and in management from Alagappa University, Karaikudi. He has got a decade's experience in academics and another seven years in industries.  He has served well known institutions including Thiagarajar School of Management and ICFAI.  His research areas of Object-oriented analysis and design and Software Engineering.  He has published five articles in reputed international journals and in many conference proceedings.  He is currently working as the Professor and HOD in the MCA department of New Horizon College of Engineering, Bangalore.

**Padmavathi Ganapathi** is the Professor and Head, Department of Computer Science of Avinashilingam University for Women, Coimbatore, India. She has 50 publications at National and International level. Her research interests are Computer Networks and Genetic Algorithms.  Contact her at ganapathi.padmavathi@gmail.com.

**Iyakutti Kombiah** is a Senior Professor of School of Physics of Madurai Kamaraj University, Madurai, India. His research interests are Computational Physics and Software Engineering.  Contact him at iyakutti@yahoo.co.in

**Mani MNS** has worked in senior positions with leading corporate houses and has a overall experience of two decades.  He is presently the consultant, Lakshmi Systems, Madurai.  Contact him at mnsmani@gmail.com.